\documentclass[12pt,a4paper]{article}
\usepackage{latexsym}
\usepackage{amsfonts,amsbsy,amssymb,epsfig}
\usepackage{graphics,graphicx}
\usepackage{amsmath}
\usepackage{graphicx}
\usepackage{color}

\begin{document}

\begin{center}

\noindent {\Large \bf Analytic solutions for the $\Lambda$-FRW
Model} \vskip 2.0cm \noindent {{\bf R. Aldrovandi}$^{1}$,  {\bf R.R.
Cuzinatto}$^{2}$ and {\bf L.G. Medeiros}$^{3}$}

\vskip 1.0cm

{\it Instituto de F\'{\i}sica Te\'orica, S\~ao Paulo State University\\
Rua Pamplona 145, 01405-900 S\~ao Paulo SP, Brazil}

\vskip 1.0cm

%%%%%%%%%%%%%%%%%%%%%%%%%%%%%%%%%%%%%%%%%%%%%%%%%%%%%%%%%%%%%%%%%%%%%
%%%%%%%%% the e-mails are in order of the author above %%%%%%%%%%%%%%
%%%%%%%%%%%%%%%%%%%%%%%%%%%%%%%%%%%%%%%%%%%%%%%%%%%%%%%%%%%%%%%%%%%%%
$^{1}$ ra@ift.unesp.br, $^{2}$ rodrigo@ift.unesp.br, $^{3}$ leo@ift.unesp.br \\
%%%%%%%%%%%%%%%%%%%%%%%%%%%%%%%%%%%%%%%%%%%%%%%%%%%%%%%%%%%%%%%%%%%%%

%\date{\today}

\end{center}

\vskip 2.0cm

%\begin{abstract}

{\it \noindent The high precision attained by cosmological data in
the last few years has increased the interest in exact solutions.
Analytic expressions for solutions in the Standard Model are
presented here for all combinations of $\Lambda = 0$ , $\Lambda \ne
0$, ${\kappa} = 0$ and ${\kappa} \ne 0$, in the presence and absence
of radiation and nonrelativistic matter.  The most complete case
(here called the ${\Lambda}{\gamma}CDM$ Model) has $\Lambda \ne 0,
{\kappa} \ne 0$, and supposes the presence of radiation and dust. It
exhibits clearly the recent onset of  acceleration. The treatment
includes particular models of interest such as the $\Lambda$CDM
Model (which includes the cosmological constant plus cold dark
matter as source constituents).}

 \vskip 0.5cm \noindent {\bf Running head}:
{$\Lambda$-FRW Model}

%\end{abstract}

\vfill \eject

%%%%%%%%%%%%%%%%%%%%%%%%%%%%%%%%%%%%%%%%
%%%%%%%%%%%% Introduction %%%%%%%%%%%%%%
%%%%%%%%%%%%%%%%%%%%%%%%%%%%%%%%%%%%%%%%

\section{{\bf INTRODUCTION}}

Recent cosmological data~\cite{novae1,BBR1,WMAP} suggest a
present-day Universe dominated by dark energy, though with a
significant contribution of matter (visible and dark).  This brings
to the fore a Friedmann-Robertson-Walker model with a strong de
Sitter component, that is, with a cosmological constant.  In the
usual approach to standard cosmology the history of the Universe is
divided into eras, each one dominated by one of the
source-constituents.  Separate solutions are used for each era.  The
intermediate periods were described by approximations which were
quite acceptable until recently.  The precision of present-day data,
however, makes it desirable to have solutions as complete and exact
as possible.  An exact solution for the Standard Model can in
principle be obtained provided an equation of state (giving the
pressure in terms of the energy density) is given for each
constituent.  For all times except the first few seconds (the
equation of state for a very-very-high-temperature black-body is
unknown), the main difficulty concerns dark matter.  It is our aim
here to  provide a short,  compact presentation, in terms of
present-day parameters, of the  solutions \cite{Co82,Da85,Da96} for
a dust dark matter, which can be called ``${\Lambda}{\gamma}CDM$
Model''. This  complete,  unified version is a necessity  whenever
applications and comparisons are in view.  We shall, however,
proceed step by step, the reason for that being that the general,
complete solution is actually  implicit (gives time as a function of
the scale parameter) and involves elliptic functions of the first
and the third kinds.  Such a progressive presentation seems
advisable, describing solutions for particular values of the
cosmological constant and curvature parameters, presence and absence
of matter and radiation, with increasing complexity.

Section \ref{sec:Generaloutlook} is a short introduction to the
Friedmann equations \cite{Nar93,Wei72,LL75,CE97,ZN71}, the main
objective being the introduction of convenient parameters and
notation.  Maybe too much detail on well-known results is given in
section \ref{sec:dSinflation}, concerned with de Sitter solutions.
The assumptions there made concerning inflation will, however,
propagate to the ensuing solutions.  In effect, we use the standard
exponential inflation there found as a kind of ``correspondence
principle'': all other solutions will generalize that case, and
reduce to it for the corresponding choice of parameters. Section
\ref{sec:gammadS} treats the model with radiation, cosmological
constant and curvature.  The limits to particular cases, as the pure
radiation model, are readily obtained.  In section
\ref{sec:BeyonddSmatter} we analyze the case for matter plus another
contribution, either the cosmological constant or curvature.  Adding
any other contribution leads to a solution as involved as the
general case, which is given in Section \ref{sec:GeneralSolution}.

%%%%%%%%%%%%%%%%%%%%%%%%%%%%%%%%%%%%%%
%%%%%%%%% General outlook %%%%%%%%%%%%
%%%%%%%%%%%%%%%%%%%%%%%%%%%%%%%%%%%%%%

\section{{\bf GENERAL OUTLOOK}}
\label{sec:Generaloutlook}

Let us start fixing notation by recalling Einstein's equations with a
cosmological--constant and source energy content modeled by a perfect
fluid:
\begin{gather}
\notag R_{\mu \nu }-{\textstyle\frac{1}{2}}R \ g_{\mu \nu } -
\Lambda g_{\mu \nu}=\frac{8 \pi G}{c^{4}} \ T_{\mu \nu } ,
\label{cosmoEinstein} \\ T_{\mu\nu} = (p + \rho c^{2}) \ u_{\mu}
u_{\nu} - p \ g_{\mu\nu} \; .
 \label{Tmunu}
\end{gather}
Here $\rho$ = $\epsilon/c^{2}$ is the mass equivalent of the energy
density.  We use metric signature $(+,-,-,-)$, so that $u_{\mu}u^{\mu}
= 1$ for timelike flux lines.  Spacetimes with homogeneous and
isotropic space sections are described by the Robertson--Walker line
element~\cite{ Nar93, Wei72}
 \begin{equation}
ds^{2} = c^{2} dt^{2} - a^{2}(t) \left[ \frac{dr^{2}}{1 - {\kappa}
r^{2}} + r^{2} d\theta^{2} + r^{2} \sin^{2}\theta d \phi^{2} \right]
\label{Friedds2}
 \end{equation}
with ${\kappa} = 0, \pm 1$, which reduces Einstein's equations to the two
Friedmann equations for the scale parameter a(t):
\begin{gather}
{\dot a}^{2} = \left[ 2 \left( \frac{4 \pi G}{3} \right) \rho+
\frac{\Lambda c^{2}}{3} \right] a^{2} - {\kappa} c^{2} \; ,
\label{Friedmann1} \\
  {\ddot a} = \left[ \frac{\Lambda c^{2}}{3} - \frac{4 \pi G}{3}
  \left( \rho+ \frac{3 p}{c^{2}} \right) \right] a \; .
  \label{Friedmann2}
\end{gather}

Combining the last two equations, one finds the expression for energy
conservation,
 \begin{equation}
 \frac{d\rho}{da} = -\, \frac{3}{a} \left( \rho + \frac{p}{c^2}
 \right) .  \label{conservation}
 \end{equation}
The total matter energy density $\epsilon_{m}= \rho_m c^2= \rho_b c^2+
\rho_\gamma c^2$ includes both non-relativistic matter (``baryons''),
dark or not, and radiation.  It is convenient to relate to $\Lambda$
the ``dark energy'' density
\begin{eqnarray}
\epsilon_{\Lambda} = \frac{\Lambda c^4}{8 \pi G} \,\,\, .
\label{eq:darkenergy}
\end{eqnarray}
In terms of the Hubble function
\begin{equation}
H(t) = \frac{{\dot a}(t)}{a(t)}
\end{equation}
Eqs.(\ref{Friedmann1},\ref{Friedmann2}) assume the forms
\begin{eqnarray}
H^2 &=& \frac{8 \pi G}{3 c^2}\, \left[\epsilon_{m} +
\epsilon_{\Lambda}
\right]\,-\, \frac{{\kappa} c^2}{a^2} \, , \label{Friedmann_H1} \\
{\dot{H}} &=& \frac{{\kappa} c^2}{a^2} - \frac{3}{2}\, \frac{8 \pi
G}{3 c^2}\, \left[\epsilon_{m} + p_{m} \right] .  \label{Friedmann_H2}
\end{eqnarray}
Actually, these equations acquire their most convenient form in terms
of dimensionless variables, obtained by dividing by the present-day
value $H^2(t=t_0) = H_0^2$ and introducing the parameters
\begin{eqnarray}
\Omega_{b} (t) = \frac{\rho_b (t) }{\rho _{crit}} \quad &;& \quad
\Omega _{m} = \Omega _{b} +\Omega _{\gamma} = \frac{\rho_b}{\rho
_{crit}} + \frac{\rho_\gamma}{\rho _{crit}} \, ,\\
\Omega_{{\kappa}}(t) = -\, \frac{{\kappa}c^{2}}{a^{2}H_{0}^{2}} \quad
&;& \quad \Omega_{\Lambda } = \frac{\Lambda c^{2}}{3H_{0}^{2}} =
\frac{\rho_{\Lambda}}{\rho_{crit}} \; ,
\end{eqnarray}
with $\rho_{crit} = \frac{3 H_0^2}{8 \pi G}$.  The first equation, in particular,
becomes
\begin{equation}
\frac{H^{2}}{H_{0}^{2}}=\frac{\rho_m}{\rho _{crit}}-\frac{{\kappa}
c^{2}}{ a^{2}H_{0}^{2}}+\frac{\Lambda c^{2}}{3H_{0}^{2}} \equiv \Omega
_{m}(t) + \Omega _{{\kappa}}(t) + \Omega _{\Lambda }, \label{H2again}
\end{equation}
which leads to the usual normalization for present-day values,
\begin{equation}
\Omega_{m0} + \Omega_{{\kappa}0} + \Omega_{\Lambda } = 1,
\label{constraint1}
\end{equation}
or
\begin{equation}
\left( \Omega_{m0} + \Omega_{\Lambda } \right) - 1 = \Omega_{total} -
1 = \frac{{\kappa} c^{2}}{ a_0^{2}H_{0}^{2}}\,\,\, .
\end{equation}
Observation will fix the value of ${\kappa}$: we are forced to choose
${\kappa} = +1$ if $\Omega_{total} > 1$ and ${\kappa} = -1$ if
$\Omega_{total} < 1$.  In both cases the present-day value of the
scale factor is determined by $a_0 =
\sqrt{\frac{{\kappa}}{\Omega_{total}-1}}\, \frac{c}{H_0} $.  If
$\Omega_{total} = 1$, then ${\kappa} = 0$ and $a_0$ remains
undetermined.

Pressure in equation (\ref{conservation}) is the sum $p = p_m = p_b +
p_\gamma$, the necessary expressions being given by thermodynamic
considerations.  Non-relativistic matter is represented as an ideal
gas, i.e., $p_b = nkT$.  This pressure, however, is negligible against
its related density (dust approximation): $(\rho_b + p_b/c^2) =
\frac{n}{c^2} (mc^2 + kT) \simeq \rho_b$, since $mc^2 \gg kT$.  The
radiation equation of state is that of a black body~\cite{LandauLifL74},
$p_\gamma = \frac{\rho_\gamma c^2}{3} = \frac{ \epsilon}{3} \,\, $.

The conservation equation (\ref{conservation}), which is
\begin{equation}
  \left[ \frac{d\rho_b}{da} + 3 \frac{\rho_b}{a} \right] + \left[
  \frac{d\rho_\gamma}{da} + 4 \frac{\rho_\gamma}{a} \right] = 0\, ,
   \label{b_gamma}
 \end{equation}
can be solved immediately under the hypothesis of independence of the
components: matter and radiation have their own separated dynamics
with respect to the scale factor $a$.  The result is the pair
\begin{eqnarray}
  \rho_b = \rho_{0b}\left(\frac{a_0}{a}\right)^3 =
  \rho_{0b}\left(1+z\right)^3 \, &;& \, \rho_\gamma = \rho_{0\gamma}
  \left(\frac{a_0}{a}\right)^4= \rho_{0\gamma} \left(1+z\right)^4
  \label{rho_b_gamma}
 \end{eqnarray}
where $z$ is the red-shift and $\rho_0 = \rho(a_0)$. These results
for the relationship between density and scale factor for matter and
radiation is a natural consequence of the baryonic number
conservation, which is strongly supported by the very high value of
the proton lifetime~\cite{ParticleData}.  In fact, if the number of baryons
$N_b = n_b a^3$ is set constant, $\frac{dN_b}{dt}=0$, it follows that
$n_b a^3 = n_{b0} a_0^3$, $n_b$ being the numerical baryonic density.
The first of equations (\ref{rho_b_gamma}) then follows because
$\rho_b = n_b (m_b c^2)$, and leads to the second once inserted into
(\ref{b_gamma}).

Finally, we introduce the notations
\begin{gather}
 \gamma c^2 = \frac{8 \pi G}{3} \rho_{0\gamma} a_0^4 = \Omega_{\gamma
 0} H_0^2 a_0^4 \label{eq:defparamGamma} \, ,\\
 M c^2 = \frac{8 \pi G}{3} \rho_{0b} a_0^3 = \Omega_{b0} H_0^2 a_0^3
 \label{eq:defparammatter} \, ,\\
 \frac{c^2}{L^2} = \frac{\Lambda c^2}{3} = \Omega_\Lambda H_0^2
 \,,\label{eq:defparamL}
\end{gather}
in terms of which equations (\ref{Friedmann1},\ref{Friedmann2}) take
on the simple forms we shall be using:
\begin{gather}
\frac{{\dot a}^2}{a^2} = \frac{c^2}{L^2} + \frac{\gamma c^2}{a^{4}} +
\frac{M c^2}{a^{3}} - \frac{{\kappa} c^2}{a^2} \, ; \label{solsimple1}\\
 \frac{{\ddot a}}{a} = \frac{c^2}{L^2} - \frac{\gamma c^2}{{a}^4} -
 \frac{M c^2}{2\,{a}^3}\, .
 \label{solsimple2}
\end{gather}

In proceeding to our step-by-step approach, it will be convenient to
start from the simplest and dominant solution, de Sitter spacetimes.
They will help fixing some aspects to be retained in the other cases.
Notice that (\ref{solsimple1}, \ref{solsimple2}) impose severe
conditions on parameters for the particular solutions we shall be
examining.  An example which we shall bypass is the case $\Lambda =
\gamma = M = 0$, which would only be admissible for $\kappa = 0, -\, 1$.

%%%%%%%%%%%%%%%%%%%%%%%%%%%%%%%%%%%%%%%%%
%%%%%%%% de Sitter solutions %%%%%%%%%%%%
%%%%%%%%%%%%%%%%%%%%%%%%%%%%%%%%%%%%%%%%%

\section{{\bf DE SITTER SOLUTIONS}}
\label{sec:dSinflation}

We recall that there are two different kinds of such spaces:
\begin{enumerate}
\item {de Sitter spacetime proper}: indicated dS: a one-sheeted
hyperbolic 4-space with topology $R^1 \times S^3$, an inclusion in the
pseudo--Euclidean space ${\bf E}^{4,1}$ satisfying ($A, B = 0, 1, 2,
3, 4$)
 \begin{equation}
\eta_{A B} \, \xi^A \xi^B = \eta_{\alpha \beta} \, \xi^{\alpha}
\xi^{\beta} - \left(\xi^4\right)^2 = -\, {L}^2
\label{deSitterspace}
\end{equation}
Its scalar curvature $R$ is~---~within our conventions~---~negative.

\item {anti--de Sitter spacetime}: indicated AdS: a two-sheeted
hyperbolic 4-space with topology $S^1 \times R^3$, an inclusion in the
pseudo--Euclidean space ${\bf E}^{3,2}$ satisfying
\begin{equation}
\eta_{A B} \, \xi^A \xi^B = \eta_{\alpha \beta} \, \xi^{\alpha}
\xi^{\beta} + \left(\xi^4\right)^2 = {L}^2 \; .
\label{antideSitterspace}
\end{equation}
\end{enumerate}
It has $R > 0$.  With the notation $\eta_{44} = s$, dS and AdS can be
put together in
\begin{equation}
\eta_{A B} \, \xi^A \xi^B = \eta_{\alpha \beta} \, \xi^{\alpha}
\xi^{\beta} + s \left(\xi^4\right)^2 = s {L}^2 \; .
\label{DSbothdeSitter}
\end{equation}
Both dS ($s = - 1, \Lambda > 0$) and AdS ($s = + 1, \Lambda < 0$) are
solutions of the sourceless Einstein's equations with a cosmological
constant related to the pseudo-radius $L$ and the scalar curvature by
\begin{equation}
\Lambda = -\, s\, \frac{3}{L^2} = -\, \frac{R}{4} \,\, .
\label{deSitterLambda}
\end{equation}
As they are sourceless solutions and have homogeneous and isotropic
space sections, these spacetimes satisfy equations (\ref{solsimple1})
and (\ref{solsimple2}) with vanishing $\gamma$ and $M$:
\begin{gather}
{\dot a}^{2} = -\, s \, \left(\frac{c}{L} \right)^2 a^2 - {\kappa}
c^{2} \; ; \label{cosmoDS1} \\
{\ddot a} = -\, s \, \left( \frac{c}{L} \right)^2 a\,\, .
\label{cosmoDS2}
\end{gather}

Let us start with case ${\kappa} = 0$.  The second equation has
solutions of the general form $a (t) = A_0 e^{\alpha t} + B_0 e^{\beta
t}$, actually with $\,\,\alpha$ = $\beta$.  Thus, the general
expression reduces to the form $a (t)$ = A $e^{\alpha t}$.  But then
${\dot a}/a$ = $H$ = $\alpha$ = $ \pm \sqrt{-s}\,\frac{c}{L}$.  The
only way to have a real positive constant $H$ is to choose $s <0$ and
the upper sign.  These choices lead to
\begin{equation}
 a (t) = A \, e^{\frac{c}{L} t} \, ,
  \label{cosmoDSsolk=0}
\end{equation}
the standard inflationary solution of exponential type.  The initial
condition is set for the beginning of the time measure, i.e., $A =
a(t=0)$.  We could equivalently choose as ``initial'' value the
present time $t_0$, such that $A=a_0 e^{-\, \frac{c}{L} t_0}$.

Notice that $\Lambda = -\, s \frac{3}{L^2} > 0$ is essential for
inflation~---~only the dS space (and not the AdS) can lead to
inflation.  Another point: the usual treatment considers sometimes
solutions with initial value $A = a(0) = 0$.  Here, however, there
will be no inflation if the initial value of $a(t)$ is zero.  We shall
from now on consider only $A > 0, \Lambda > 0$, and look for solutions
generalizing this result, that is, which reduce to
(\ref{cosmoDSsolk=0}) when only a cosmological term is present and
${\kappa} = 0$.

For ${\kappa} \ne 0$,§ the solution allowing for inflation is
\begin{eqnarray}
a (t) &=& A \cosh {\textstyle{\frac{ct}{L}}} + \sqrt{A^2 - {\kappa} {L}^2}
\sinh {\textstyle{\frac{ct}{L}}} = \nonumber \\
 &=& \frac{1}{2} \left[ \left( A + \sqrt{A^2 - {\kappa}L^2} \right)
 e^{\frac{ct}{L}} + \left(A - \sqrt{A^2 - {\kappa}L^2} \right)
 e^{-\frac{ct}{L}} \right] .
 \label{eq:dSsolutiongeneral6}
\end{eqnarray}
If we introduce the
function
\begin{equation}
f(t) = \left( A + {\sqrt{A^2 - {\kappa}\,{L}^2}} \right)\,
e^{\frac{c\,t}{{L}}} \label{eq:dSfunctionf}
\end{equation}
and use the relation
\begin{equation}
\left( A - \sqrt{A^2 - {\kappa}L^2} \right) =
\frac{{\kappa}L^2}{\left( A + \sqrt{A^2 - {\kappa}L^2} \right)}
\quad \quad ({\kappa} \ne 0) \label{eq:dSrelation} \, ,
\end{equation}
solution (\ref{eq:dSsolutiongeneral6}) takes on the form
\begin{equation}
a (t) =  \frac{1}{2} \left[f(t) + \frac{{\kappa}L^2}{f(t)} \right] .
\label{eq:dSsolutiongeneral7}
\end{equation}

Summing up, the solution is
\begin{equation}
\Lambda =  \frac{3}{L^2} \ne 0: \left\{
\begin{matrix}
a (t) = \frac{1}{2} \left[ f(t) + \frac{{\kappa}{L}^2}{f(t)} \right]
\quad {\mbox{with}}\\
f(t) = \left( A + {\sqrt{A^2 - {\kappa}{L}^2}} \right)
e^{\frac{ct}{L}}
\end{matrix}
\right.
\end{equation}
and the Friedmann-Robertson-Walker form of the inflationary de Sitter
line element will consequently be
\begin{equation}
ds^2 = c^2 dt^2 - \frac{1}{4} \left[ f(t) +
\frac{{\kappa}{L}^2}{f(t)} \right]^2 \left[
\frac{dr^2}{1-{\kappa}r^2} + r^2 (d\theta^2 + \sin^2\theta d\phi^2)
\right] \, .
\end{equation}

%%%%%%%%%%%%%%%%%%%%%%%%%%%%%%%%%%%%%%%%%%%%%%%%%%%%%%%%%%%%
%%%%%%%%%%%%%%%%%%%%%% Radiation %%%%%%%%%%%%%%%%%%%%%%%%%%%
%%%%%%%%%%%%%%%%%%%%%%%%%%%%%%%%%%%%%%%%%%%%%%%%%%%%%%%%%%%%

\section{{\bf RADIATION WITH ${\mathbf{{\kappa} \ne 0, \Lambda \ne 0}}$}}
\label{sec:gammadS}

From all the evidence we have today, the Universe has passed an
initial stage during which, besides the cosmological term,
ultrarelativistic matter or radiation provided a significant
contribution as a source in Einstein's equations.  This era is
described by equations (\ref{solsimple1}) and (\ref{solsimple2}) with
$M = 0$,
\begin{gather}
 {\dot a}^2 = \left[ \frac{\gamma c^2}{a^{4}} + \frac{c^2}{L^2}
 \right] a^2 - {\kappa} c^2 \, ; \label{Fried_rad1} \\
  {\ddot a} = \left[ \frac{c^2}{L^2} - \frac{\gamma c^2}{{a}^4}
  \right] a \, .
 \label{Fried_rad2}
\end{gather}

We solve the first of these equations by direct integration
(completing squares and changing variables) with our conventional
initial condition $a(0)=A$.  The result reads:
\begin{equation}
  a(t)= \sqrt{ \frac{{\kappa}L^2}{2} - \frac{1}{2} \left[ g(t) +
  \frac{ \left( \frac{{\kappa}L^2}{2} \right)^2 - \gamma
  L^2}{g(t)}\right]}
 \label{a_rad}
\end{equation}
with the definition
\begin{eqnarray}
 g(t) \equiv \left[ \left( \frac{{\kappa}L^2}{2} - A^2 \right) -
 \sqrt{\gamma L^2 + A^2 (A^2 - {\kappa}L^2)} \right] e^{2 \frac{c}{L}
 t} \label{g} .
\end{eqnarray}

The solution above involves, besides the cosmological constant and
${\kappa}$, other quantities hidden in parameter $\gamma$: the
present-day value $a_0$ of the expansion parameter and the actual
value of the radiation density $\rho_{\gamma 0}$.  At $t = 0$ we
recover the initial condition $a(0) = A$, as it should be.

It is a good test of consistency to analyze some limits.  \textbf{(i)}
In the absence of radiation, $\gamma = 0$, solution (\ref{a_rad})
reduces dutifully to the de Sitter solution for ${\kappa} \ne 0$,
equation (\ref{eq:dSsolutiongeneral6}).  \textbf{(ii)} In the limit of
no cosmological constant, $L \rightarrow +\infty $, the scale factor
(\ref{a_rad}) is simplified to
\begin{equation}
a(t) = \sqrt{A^2 + 2\sqrt{\gamma -{\kappa}A^2} \, ct -{\kappa}(ct)^2}.
\label{a_rad_Lambda0}
\end{equation}
\textbf{(iii)} Taking only ${\kappa} = 0$ in (\ref{a_rad}), we are
left with:
\begin{eqnarray}
 a(t) = \sqrt{\frac{1}{2} \left\{ \left[ A^2 + \sqrt{\gamma L^2 + A^4}
 \right] e^{2 \frac{c}{L} t} + \left[ A^2 - \sqrt{\gamma L^2 + A^4}
 \right] e^{-2 \frac{c}{L} t} \right\}}.
 \label{a_rad_k0}
\end{eqnarray}
\textbf{(iv)} If we, in addiction the restriction ${\kappa} = 0$,
perform the limit $\Lambda \rightarrow 0$ (i.e. $L \rightarrow
\infty$), it follows
\begin{equation}
 a(t) = \sqrt{A^2 + 2 \sqrt{\gamma} \, ct}\, ,
 \label{a_rad_k0_Lambda0}
\end{equation}
the conventional solution for a radiation dominated universe, in which
$a \propto t^{1/2}$.

%%%%%%%%%%%%%%%%%%%%%%%%%%%%%%%%%%%%%%%%%%%%%%%%%%%%%
%%%%%%%%%%%%%% Non-relativistic matter %%%%%%%%%%%%%%
%%%%%%%%%%%%%%%%%%%%%%%%%%%%%%%%%%%%%%%%%%%%%%%%%%%%%

\section{{\bf NON-RELATIVISTIC MATTER}}
\label{sec:BeyonddSmatter}

This section deals with the possible cosmological models in absence of
radiation: $\gamma = 0$.  It is divided in two special cases: first,
besides $M \ne 0$, one assumes $\Lambda \ne 0$ but ${\kappa} =0$; and,
then, one considers no cosmological constant, $\Lambda = 0$, but
${\kappa}\ne 0$.

%%%%%%%%%%%%%%%%%%%%%%%%%%%%%%%%%%%%%%%%
%%%%%%%%%%% Lambda CDM %%%%%%%%%%%%%%%%%
%%%%%%%%%%%%%%%%%%%%%%%%%%%%%%%%%%%%%%%%

\subsection{Matter with ${\mathbf{\Lambda \ne 0, {\kappa} = 0}}$
(${\mathbf{\Lambda CDM}}$ Model)}

Equation (\ref{Friedmann_H1}) can be used to eliminate the term
$\epsilon _{m}$ from (\ref{Friedmann_H2}).  The dust approximation
$p=0$ leads then, for ${\kappa}=0$, to
\begin{equation}
\dot{H}=\frac{3}{2}\left( \frac{c^{2}}{L^{2}}-H^{2}\right) \, ,
\label{diff_eq_H}
\end{equation}
whose integration is straightforward:
\begin{equation}
H\left( t\right) =\frac{c}{L}\frac{\left[ \frac{c}{L}+H\left( 0\right) %
\right] e^{3\frac{c}{L}t}\mp \left[ \frac{c}{L}-H\left( 0\right) \right] }{%
\left[ \frac{c}{L}+H\left( 0\right) \right] e^{3\frac{c}{L}t}\pm
\left[ \frac{c}{L}-H\left( 0\right) \right] }\,\, .  \label{H Materia
Lambda}
\end{equation}
The integration constant $H\left(0\right) $ is obtained from
(\ref{solsimple1}) in terms of previously defined measurable
quantities:
\begin{equation}
H^{2}\left( 0\right) =\frac{Mc^{2}}{A^{3}}+\frac{c^{2}}{L^{2}}\, \, .
\label{H Materia Lambda inicial}
\end{equation}
The expansion parameter follows from expression (\ref{solsimple1}) for
$H^{2}$:
\[
a(t) = \left( Mc^{2}\right) ^{\frac{1}{3}}\left[ H^{2}(t)
-\frac{c^{2}}{L^{2}}\right] ^{-\frac{1}{3}},
\]
or, substituting (\ref{H Materia Lambda}) with the choice of the upper
sign,
\begin{equation}
a\left( t\right) =\frac{A}{2^{\frac{2}{3}}}\left( \left[
1+\sqrt{1+\frac{ ML^{2}}{A^{3}}}\right]
e^{\frac{3}{2}\frac{c}{L}t}+\left[ 1-\sqrt{1+\frac{
ML^{2}}{A^{3}}}\right] e^{-\frac{3}{2}\frac{c}{L}t}\right)
^{\frac{2}{3}}.  \label{a Materia Lambda}
\end{equation}
The choice of sign is adequate, as it leads to the appropriate limits:
\textbf{(i)} $ M\rightarrow 0$ reduces (\ref{a Materia Lambda}) to
(\ref{cosmoDSsolk=0}); and \textbf{(ii)} with $L\rightarrow \infty $\
(i.e. $\Lambda \rightarrow 0$) we find
\[
a\left( t\right) = \left( A^{3/2}+\frac{3}{2}\sqrt{M}ct\right)
^{\frac{2 }{3}},
\]
the solution for a matter dominated universe.
In terms of still another convenient function,
\begin{equation}
j\left( t\right) = {\textstyle{\frac{1}{2}}}\, \left[
1+\sqrt{1+\frac{ML^{2}}{A^{3}}}\right] e^{\frac{3}{2 }\frac{c}{L}t},
\label{j}
\end{equation}
solution (\ref{a Materia Lambda}) becomes
\begin{equation}
a\left( t\right) = A\, \left( j\left( t\right)
-\frac{ML^{2}}{2 A^{3}}\frac{1}{j\left( t\right) }\right)
^{\frac{2}{3}}
\label{a j}\, .
\end{equation}

%%%%%%%%%%%%%%%%%%%%%%%%%%%%%%%%%%%%%%%%
%%%%%%%%%%% Matter and kappa %%%%%%%%%%%
%%%%%%%%%%%%%%%%%%%%%%%%%%%%%%%%%%%%%%%%

\subsection{Matter with ${\mathbf{\Lambda = 0, {\kappa} \ne 0}}$ }

Equations (\ref{solsimple1}) and (\ref{solsimple2}) are, in this case,
\begin{gather}
{\dot a}^2 + {\kappa} c^2 - \frac{M c^2 }{a} = 0 \\
{\ddot a} + \frac{M c^2}{2\,{a}^2} = 0 .
\end{gather}
 The first equation,
\begin{gather}
 \int \frac{a da}{\sqrt{M a - {\kappa} a^2} } = c \int dt\, ,
\end{gather}
is better solved separately for each value of ${\kappa}$:

\textbf{(I) Case $\mathbf{{\kappa} = 0}$} can be used as a test:
integration gives $a^{3/2} - A^{3/2} = \frac{3}{2} \sqrt{M} c t$, just
the result of the previous section.\\

\textbf{(II) Case} $\mathbf{{\kappa} > 0}$, actually $\mathbf{{\kappa}
= 1}$: a change of variable puts the integrals into the form
\begin{gather}
 \int_A^a \frac{\sqrt{y}\, dy}{\sqrt{1 - \frac{y}{M}} } =
\sqrt{M} c \int_0^t dt .
\end{gather}
Integration gives $a(t)$ in an implicit way:
\begin{multline}
  c t = \sqrt{M A -  A^2} - \sqrt{M a -  a^2} + \\
 M \left[\arcsin\left(\sqrt{\frac{a}{M}} \right) -
 \arcsin\left(\sqrt{\frac{A}{M}} \right) \right] .
\end{multline}

\textbf{(III) Case} $\mathbf{{\kappa} < 0}$, actually
$\mathbf{{\kappa} = -\, 1}$: integration now gives
\begin{gather}
 c t = \sqrt{M a + a^2} - \sqrt{M A + A^2} + M \ln \left[
 \frac{\sqrt{A} +\sqrt{A + M}}{\sqrt{a} +\sqrt{a + M}} \right] .
\end{gather}
When $M > 0$, a function of $\,\,\frac{a(t)}{M}\,$ turns up:
\begin{gather}
\frac{e^{\sqrt{\frac{a}{M} +
\left(\frac{a}{M}\right)^2}}}{\sqrt{\frac{a}{M}} +\sqrt{1 +
\frac{a}{M}}} = \frac{e^{\sqrt{\frac{A}{M} +
\left(\frac{A}{M}\right)^2}} }{\sqrt{\frac{A}{M}} +\sqrt{1 +
\frac{A}{M}} } e^{\frac{c t}{M}} \,\, .
\end{gather}
This can be rewritten as
\begin{gather}
\begin{cases}
F(x) = \frac{e^{\sqrt{x}\,\sqrt{1 + x}} }{\sqrt{x} +\sqrt{1 + x} }
\label{eq:Fofx}\\ \\
F\left[\frac{a(t)}{M}\right] = F\left[\frac{A}{M}\right]\, e^{\frac{c
t}{M}} \,\, .
\end{cases}
\end{gather}
Function $F(x)$ has actually a very simple, monotonic behavior, and so
has $a(t)$.

We have in this section solved Friedmann's equations with a dust
source ($M \ne 0$), choosing solutions of inflationary type.
Actually, only the cases with {\it either} $\Lambda \ne 0$ {\it or}\,
${\kappa} \ne 0$ have been presented.  The case with both $\Lambda \ne
0$ {\it and}\, ${\kappa} \ne 0$ is much more involved.  In fact, all
the remaining classes of solutions are practically as involved as the
general case including all the four components $\Lambda, {\kappa}, M$
and $\gamma$.  Let us then proceed directly to that case.

%%%%%%%%%%%%%%%%%%%%%%%%%%%%%%%%%%%%%%%%%%%%%%
%%%%%%%%%%%%%% General Solution %%%%%%%%%%%%%%
%%%%%%%%%%%%%%%%%%%%%%%%%%%%%%%%%%%%%%%%%%%%%%

\section{{\bf GENERAL CASE:
${\mathbf{\gamma  \ne 0, M  \ne 0, \Lambda \ne 0, {\kappa} \ne
0}}$}} \label{sec:GeneralSolution}

This ${\Lambda}{\gamma}CDM$ case is described by the complete
equations (\ref{solsimple1},\ref{solsimple2}).  We repeat one of then:
\begin{gather}
{{\dot a}(t)}^2 + {\kappa} c^2 - \frac{\gamma c^2}{a(t)^2} - \frac{M
c^2}{a(t)} - \frac{{a(t)}^2 c^2}{L^2} = 0 \, .
\end{gather}
We shall integrate this first-order equation, which
is equivalent to
\begin{equation}
a(t)\, {\dot a}(t) = \sqrt{ \frac{c^2\,{a(t)}^4}{L^2} + \gamma c^2 +
M c^2 a(t) - {\kappa} c^2 a(t)^2}
\end{equation}
or
\begin{equation}
\frac{c\, }{L} \int_0^t dt = \int_A^a \frac{a\, da}{\sqrt{a^4 +
\gamma L^2 + M L^2 a - {\kappa} L^2 a^2}}.
\end{equation}

The procedure to arrive at the solution is long and rather cumbersome.
Let us proceed step by step:
\begin{enumerate}
\item Rewrite the integral above in terms of the roots $\{r_i\}$ of
the denominator: \begin{gather} \notag \int \frac{a\, da}{\sqrt{a^4
+ \gamma L^2 + M L^2 a - {\kappa} L^2 a^2}} = \int \frac{a\,
da}{\sqrt{(a - r_1)(a - r_2)(a - r_3)(a - r_4)}}.
\end{gather}

\item Introduce some constants, in terms of which the roots will be
expressed later on:
\begin{eqnarray}
\notag W &=& {\left[ 27 (M L^2)^2 -2({\kappa} L^2)^3 + 72\gamma
{\kappa} L^4\right]^2} \, , \\
\notag U &=& W + \sqrt{ W^2 -4 \left[({\kappa} L^2)^2 + 12\gamma
L^2\right]^3} \,  \\ \notag V &=& \frac{1}{3} \left\{ 2{\kappa} L^2
- \left( \frac{U}{2} \right)^\frac{1}{3} - \left( \frac{2}{U}
\right)^\frac{1}{3} \left[ ({\kappa} L^2)^2 + 12\gamma \,L^2 \right]
\right\} \, , \\ X &=& (2{\kappa} L^2 - V) \, .
\end{eqnarray}

\item  The roots will then be:
\begin{eqnarray}
\notag {\mbox{\footnotesize $r_1 = \frac{1}{2} \left( \sqrt{V} -
\sqrt{X - \frac{2 M L^2}{\sqrt{V}}}\right) $}} \,\,\, , \,\,\,
{\mbox{\footnotesize $r_2 = \frac{1}{2} \left( \sqrt{V} + \sqrt{X -
\frac{2 M L^2}{\sqrt{V}}}\right) $}} \, , \\ {\mbox{\footnotesize
$r_3 = -\frac{1}{2} \left( \sqrt{V} + \sqrt{X + \frac{2 M
L^2}{\sqrt{V}}}\right) $}} \,\,\, , \,\,\, {\mbox{\footnotesize $r_4
= -\frac{1}{2} \left( \sqrt{V} - \sqrt{X + \frac{2 M
L^2}{\sqrt{V}}}\right) $}} \, .
\end{eqnarray}

\item Solutions will be implicit, and involve elliptic functions of
two kinds~\cite{GR80,AS68,EDM2}:
\begin{itemize}
\item the elliptic integral of first kind with parameter $m$ and
amplitude $\phi$:
\begin{gather}
    F[\phi,m] = \int_0^\phi \frac{1}{\sqrt{1 - m
    \sin^2\theta}}\, d\theta .
\end{gather}
\item the elliptic integral of third kind with parameter $m$,
characteristic $n$ and amplitude $\phi$:
\begin{gather}
\Pi[\phi,n, m] = \int_0^\phi \frac{1}{(1 - n \sin^2\theta) \sqrt{1 -
m \sin^2\theta}}\, d\theta .
\end{gather}
\end{itemize}

\item The characteristic and the  parameter turning  up will be:
\begin{gather}
n = \left( \frac{r_2 - r_4}{r_1 - r_4} \right) \,\, ; \,\, m =
\left( \frac{r_1 - r_3}{r_2 - r_3} \right) n \,.
\end{gather}

\item The amplitude turning up will be
\begin{gather}
\phi (a) = \arcsin  \sqrt{\frac{a - r_2}{n (a - r_1)}} \, .
\end{gather}

\item The implicit solution  will then be
\begin{eqnarray}
\frac{c t}{2 L} = \frac{(r_1 - r_2) \left\{ \Pi[\phi(a),n, m] -
\Pi[\phi(A),n, m] \right\}}{\sqrt{(r_2 - r_3)(r_1 - r_4)}}+ \notag \\
-\frac{r_1 \left\{ F[\phi(a) ,m] - F[\phi(A),m] \right\}}{\sqrt{(r_2 -
r_3)(r_1 - r_4)}} \, .  \label{a_general}
\end{eqnarray}

\end{enumerate}

We see from their definitions that both elliptic functions vanish when
$\phi = 0$.  The intricacy of this solution justifies its use only in
the general case.  It does reduce to the expected particular cases:
this can be more easily verified by confronting the plots of each
particular case with the corresponding graphics constructed from the
general scale factor obtained from (\ref{a_general}).
%%%%%%%%%%%%%%%%%%%%%%%%%%%%%%%%%%%%%%%%%%%%%%%
%%%%%%%%%%%%%%%% FIGURE  1 %%%%%%%%%%%%%%%%%%%%
%%%%%%%%%%%%%%%%%%%%%%%%%%%%%%%%%%%%%%%%%%%%%%%

%\includegraphics[height=3cm, width=4.5cm]{Figure.pdf}

\begin{figure}[ht]
\begin{center}
 \includegraphics[width=5in]{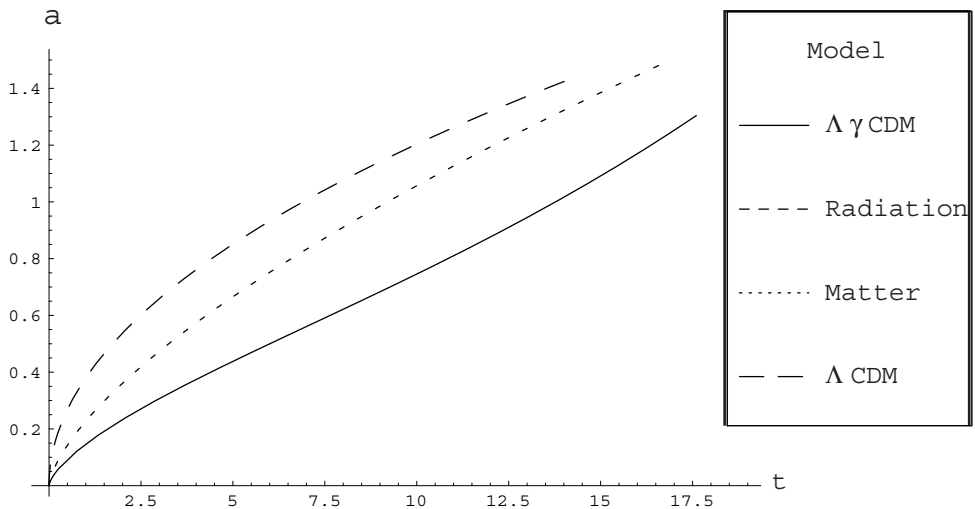}
 \caption{{\it The scale factor $a(t)/a_0$ in terms of $t$ (in
 Gyear). The two higher curves are reference models (pure radiation and
 matter, respectively) exhibiting an overall deceleration.  The full
 curve comes from the $\Lambda \gamma CDM$ universe, which coincides
 with the $\Lambda CDM$ model in the scale used.  It presents an
 inflection point, at which $\Lambda$ starts dominating and
 acceleration sets on.}}
 \label{fig1}
 \end{center}
\end{figure}
%%%%%%%%%%%%%%%%%%%%%%%%%%%%%%%%%%%%%%%%%%%%%%%

The graphs shown in Figure \ref{fig1} include the general solution
(full line, the ${\Lambda}{\gamma}CDM$ Model), and two other reference
cases: the radiation dominated Universe (dashed line) and the matter
dominated Universe (dotted curve).  Following the observational data,
the curvature parameter is set to zero for all functions $a(t)$,
${\kappa}=0$.  The values of $L$, $M$ and $\gamma$ are chosen in
accordance with the present day values \cite{WMAP,ParticleData}:
$\Omega_{\text{total}} = 1.02 \pm 0.02$, $\Omega_{b0} = 0.27 \pm 0.04$
(contribution of baryonic and dark matter) and $\Omega_{\gamma0} =
(4.9 \pm 0.5) \times 10^{-5}$, and $H_0 = (71_{-0.003}^{+0.004})$ km
sec$^{-1}$ Mpc$^{-1}$.  Time $T=ct \,\, (c=1)$ is given in Gyear\, and
the scale factor is measured in units of its present day value $a_0$,
an arbitrary constant.  When considering pure radiation (pure matter),
we take $\Omega_{\gamma0} = \Omega_{total}$ ($\Omega_{b0} =
\Omega_{total}$).

The $\Lambda \gamma CDM$ model gives $13.8$ Gyear for the age of the
universe.  Unlike the two other models, it exhibits an inflection
point for not too large a $z$, which starts the present day
acceleration.  As expected, the effect of matter and radiation
surpasses largely that of the cosmological constant in the early
Universe, and a pure radiation model evolves much faster than a matter
dominated universe.  In the present and future, however, the
cosmological constant becomes more important than radiation and matter
and dominates the cosmic dynamics.

\begin{figure}
 \begin{center}
 \includegraphics[width=5in]{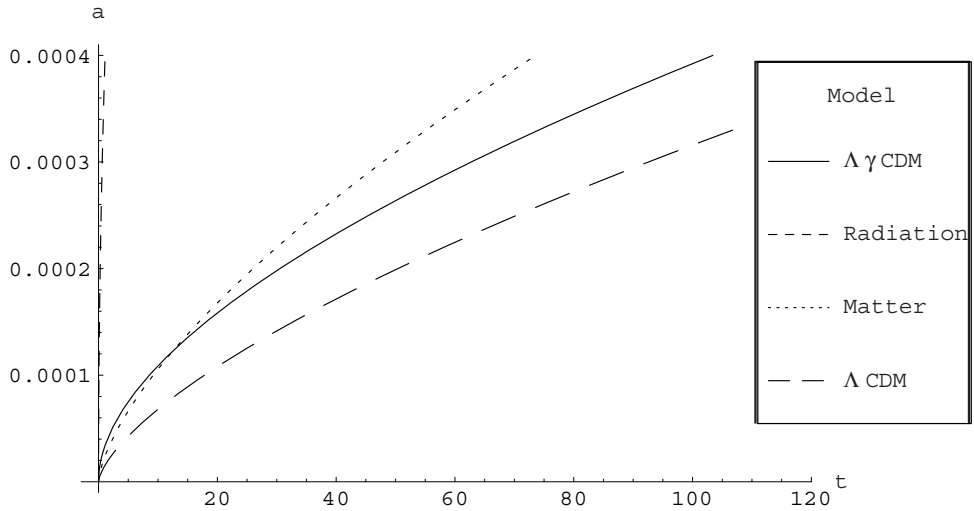}
 \caption{{\it Zooming of the curve $a(t)/a_0$ for the primeval
 universe, with time $t$ now given in $kyear$.  The dashed curve of
 the radiation model lies very close the vertical axis.  The
 difference between $\gamma \Lambda CDM$ and $\Lambda CDM$ curves
 (full line and lower dashed line) is clearly shown.} }
 \label{fig2}
 \end{center}
\end{figure}

Another interesting plot is that of matter and $\Lambda$ (with $\gamma
= {\kappa} = 0$), the $\Lambda CDM$ model.  It is in fact also plotted
in Figure \ref{fig1}, but coincides with the complete solution in the
scale used.  At small scales, however, it is clear that the general
$\Lambda \gamma CDM$ model ($\Omega_{b 0} = 0.27, \, \Omega_{\Lambda} =
0.73, \, \gamma = {\kappa} = 0$) differs from the $\Lambda CDM$ model,
which corresponds to the lower curve in Figure \ref{fig2}, which is an
enlargement of Figure \ref{fig1} for early times.  The radiation is
the dominant component in this primeval period: see the dashed curve
close to the vertical axis, or the interval ($t \lesssim 10 \, kyear$)
where complete model (full line) overcomes the matter one (dotted
line).
%%%%%%%%%%%%%%%%%%%%%%%%%%%%%%%%%%%%%%%%%%%%%%%
%%%%%%%%%%%%%%%% FIGURE  2 %%%%%%%%%%%%%%%%%%%%
%%%%%%%%%%%%%%%%%%%%%%%%%%%%%%%%%%%%%%%%%%%%%%%
%%%%%%%%%%%%%%%%%%%%%%%%%%%%%%%%%%%%%%%%%%%%%%

%%%%%%%%%%%%%%%%%%%%%%%%%%%%%%%%%%%%%%%%%%%%%%
%%%%%%%%%%%%%% Final Comments %%%%%%%%%%%%%%%%
%%%%%%%%%%%%%%%%%%%%%%%%%%%%%%%%%%%%%%%%%%%%%%

\section{{\bf FINAL COMMENTS}}

The solutions given above hold for all times, as long as dark matter
is a dust formed by baryons or, at least, when its energy density
behaves according to the pure volumetric law $\epsilon_{m} =
\epsilon_{m0}\, (1+z)^3$.  Notice that Eq.(\ref{solsimple2}) is, in
time units of $H_0^{-1}$,
\begin{gather}
\frac{{\ddot a}}{a} = \Omega_\Lambda - \Omega_{\gamma 0}\,(1+z)^4 -
{\textstyle{\frac{1}{2}}}\, \Omega_{b 0}\,(1+z)^3,
\end{gather}
so that  acceleration will change sign  when
\begin{gather}
\Omega_{\Lambda} = \Omega_{\gamma 0}\, (1+z)^{4} +
{\textstyle{\frac{1}{2}}}\, \Omega_{b 0}\, (1+z)^{3} \, .
\end{gather}
This condition is ${\kappa}$-independent.  For the present-day
favored~\cite{WMAP} values $\Omega_{\Lambda} = 0.73,\Omega_{b 0} =
0.27, \Omega_{\gamma 0} = 4.88 \times 10^{-5}$, this take-over
happens rather recently, at $z \approx 0.75$, or $a(t) \approx
0.57\, a_0$. The limitations of the $\Lambda \gamma CDM$ model can
be seen in the estimate value coming up for the cross-over, which
seems too high: $(1+z) = \Omega_{b 0}/\Omega_{\gamma 0} \approx
5530$, to be compared with the preferred WMAP value $z \approx
3234$.

We have seen that the $\Lambda CDM$ model, described by equation
(\ref{a Materia Lambda}), has a simpler form than the general $\Lambda
\gamma CDM$ solution, equation (\ref{a_general}): the latter gives the
scale factor as an implicit function of time, unlike the former.
Figure \ref{fig1} shows also that the $\Lambda CDM$ model coincides
with the complete solution at recent times.  Nevertheless, the curve
of $\Lambda \gamma CDM$ model differs from the one of $\Lambda CDM$ in
the early universe (Figure \ref{fig2}), when radiation has a dominant
role.  Whenever precision is needed, the general solution
(\ref{a_general}) is to be used.  This general solution will be valid
as long as we take the combination of ultra-relativistic and
non-relativistic ideal particles in the presence of a vacuum energy
with equation of state $p = -\,\, \epsilon$.

%%%%%%%%%%%%%%%%%%%%%%%%%%%%%%%%%%%%%%%%%
%%%%%%%%%%% Acknowledgments %%%%%%%%%%%%%
%%%%%%%%%%%%%%%%%%%%%%%%%%%%%%%%%%%%%%%%%

\section*{\small Acknowledgment}
The authors would like to thank FAPESP-Brazil and CNPq-Brazil for
financial support.

%%%%%%%%%%%%%%%%%%%%%%%%%%%%%%%%%%%%%%%%%
%%%%%%%%%%%% Bibliography %%%%%%%%%%%%%%%
%%%%%%%%%%%%%%%%%%%%%%%%%%%%%%%%%%%%%%%%%

\end{document}